\documentclass[runningheads,a4paper]{llncs}

\usepackage{enumitem}
\setlist{nolistsep}

\usepackage{amssymb}
\usepackage{graphicx}
\usepackage[usenames,dvipsnames]{color}

\usepackage{url}

\usepackage{amsmath}
\usepackage{textcomp} % \textlbrackdbl, \textrbrackdbl
\usepackage{xspace} 
\usepackage{bbding} % check marks
\usepackage{pifont} % arrows
\usepackage{stmaryrd} % arrows
\usepackage{mdwlist} %list
\usepackage[cp1252]{inputenc}

\begin{document}

\authorrunning{PREPRINT}

\title{Software Engineering Solutions\\ To Support Vertical Transportation
} 
\author{Alber J. Christianto, Peng Chen, Osheen Walawedura,\\ Annie Vuong, Jun Feng, Dong Wang,  Maria Spichkova}
\institute{RMIT University,  Australia \\  \email{$\{$s3613559,s3538152,s3573429,s3579678,s3567360,s3567724$\}$@student.rmit.edu}\\
\email{maria.spichkova@rmit.edu.au} }
\maketitle

\begin{abstract}
In this paper we introduce the core results of the project on visualisation and analysis of data collected from the vertical transport facilities. 
The aim of the project was to provide better user experience as well as  to help building maintenance staff to increase productivity of their work. 
We elaborated a web-based system for vertical transportation, to cover the needs of 
(1) staff working on building maintenance, (2) people who are regularly using the facilities in the corresponding buildings.
\end{abstract}

\section{Introduction}

Optimisation of the vertical transport facilities, such as lifts and escalators, is twofold: on the one hand the optimisation should be focused on providing the best service for the users, but on the other hand the related cost have to be minimised.  
The health factor should also be taken into account - the users have to be encouraged to use stairs if they are going few levels up or down.

Thus, the goal of our project was to elaborate a solution that provides real-time information on the status of vertical transportation facilities within a complex of buildings. 
This would help to predict and enhance footfall, assess and resolve congestions issues, as well as study behaviour to environmental changes such as signage or overcrowding. 

The elaborated solution was implemented as a web-based system for the 
New Academic Street (NAS) precinct of the RMIT University, Australia. It includes two web applications
to monitor lifts and escalators in the NAS precinct and analyse their performances. 
The first web application is focused on the needs of staff working on building maintenance: it provides detailed information about lifts in RMIT Building 8, 10, and 12, such as lift status, lift maintenance history, lift waiting time, and emergency log. 
The second application is focusing on needs of people who are regularly using the facilities in the corresponding buildings. 
the system also provides an interactive option to find the fastest way to reach a destination using three vertical transportation mode provided by RMIT: lift, escalator, or stairs.   

\emph{Outline:} The rest of the paper is organised as follows. 
Related work is discussed in Section~\ref{sec:related}. 
Section~\ref{sec:arch} introduces the architecture of the proposed system, where the core functionality is demonstrated in Section~\ref{sec:funct}. 
Finally, Section~\ref{sec:conclusions} summarises the paper and introduces directions of our future work.

%==================================================
\section{Related Work}
\label{sec:related}

The results of a project on automated maintenance inspection, performance monitoring and ride quality measurement in vertical transportation systems were presented in 
\cite{francik2016real}. This approach is the most related to the work we present in this paper: 
The approach enabled lift performance profiles to be compiled automatically and transmitted in real time. The authors demonstrated that this solution significantly rationalized and improved the process of maintenance inspection and monitoring. 

An design methodology for elevator systems using rules and graphical methods was proposed in \cite{al2013automated}. 
Evaluating the elevator round trip time for multiple entrances and incoming traffic conditions using Markov Chain Monte Carlo was 
proposed in \cite{al2014evaluating}.
Optimization of waiting and journey time in group elevator system using genetic algorithm was proposed in  \cite{tartan2014optimization}.

Analysis of possible two dimensional elevator traffic systems in large buildings was presented in \cite{so2014analysis}.  A Petri net-based modeling and control of the multi-elevator systems was discussed in \cite{ahmad2014petri}.

One of important aspects of the project we present in this paper, was to embed the corresponding development activities  into the learning and teaching activities within the School of Science, RMIT University, Australia. 
RMIT University aims to create transformative experiences for its students, so that the students are ready  for their next steps in life and work.
The 2020  RMIT strategic plan defines the goals that we will pursue together over the following years, and one of the priorities in this plan is \emph{research embedded in teaching and engagement}. 
In our earlier work we presented how collaborative industrial project are embedded into engineering curriculum in the School of Science (disciplines \emph{Computer Science and Software Engineering} and \emph{Computer Science and IT}) and the School of Engineering, \cite{clunne2017modelling,iceer_projects,spichkova2017autonomous,spichkova2016formal,spichkova2015formal}. 
These project covered  research areas of software engineering,  robotics, as well as autonomous and automotive systems.

Sedelmaier and Landes \cite{Sedelmaier2015} provided justification from a pedagogical point of view that project-based learning (PBL) allows learners not only to gain technical knowledge in the related area, but also foster soft skills. 
Dagnino \cite{Dagnino2014} presented a method derived from the collaboration between North Carolina State University and ABB. This method brings diverse techniques to simulate an industrial environment for teaching a senior level Software Engineering course.  
In contrary to the approach of Dagnino, we do not simulate the industry environment for our projects, but use a real one, based on collaboration with local and overseas industrial partners. In addition to that, project based learning is seen as a contributing factor to graduates' better work preparation, cf. \cite{jollands2012project}.

%==================================================
\section{System Architecture}
\label{sec:arch}

\begin{figure}[ht!]
\begin{center}
\includegraphics[scale=0.9]{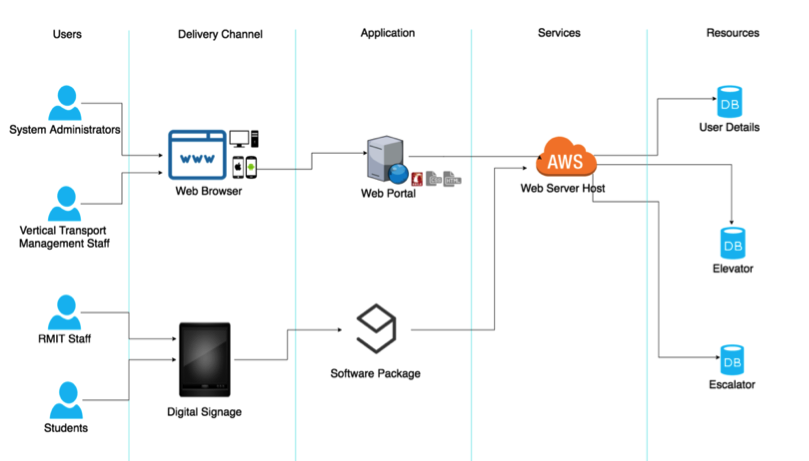}
\end{center}
\caption{System Architecture}
\label{fig:arch}
\end{figure}

The architecture of the proposed system is demonstrated in Figure~\ref{fig:arch}. 

The system deals with four types of users: 
\begin{itemize}
\item system administrators (Admins), 
\item Vertical Transportation (VT) Management staff, 
\item students, and 
\item university staff.  
\end{itemize}
Admins and VT Management staffs access the service through web browser to access the web portal, which  is a mobile-friendly web site. The portal provides different ways of displaying the information for desktops and smart phones, to adjust the presentation style to the screen size appropriately.

Accesses through web portal is divided into two parts: public and authorised users. For the public access, anyone who knows the web portal URL can access the dashboard and see the information displayed there. For authorised users access, admins and VT Management staffs have to sign in first with their valid credentials before they can view more information on the web portal.
Students and university staffs access the service through digital signage in form of display panels, which are displayed on Level 4 on Building 10 and 12. It is available for the public, but it is intended specifically for students and staffs. The display panels show general information about lifts in their particular building. For example, if the display panel is located at Building 12, then the display panel is showing information about lift statuses in Building 12 and estimated travelling time from Level 4 to other levels in Building 12 based on the lift historical data in Building 12.

Both implementations access the web server hosted by Amazon Web Service (AWS), where AWS access the databases for lifts and users. The database itself has to stored in AWS Relational Database Service (RDS).  
 
 AWS was selected as Amazon has the best reliability among the leading cloud service providers. Based on an analysis of downtime at Infrastructure-as-a-Service (IaaS) public cloud providers done by CloudHarmony in 2015 \cite{NetworkWorld}, AWS had the least downtime with 56 outages and a total of 2 hours and 30 minutes. In comparison, Microsoft Azure experienced 71 outages with a total of 10 hours and 49 minutes downtime and Google Cloud Platform had 167 outages totalling 11 hours and 34 minutes. 
 This solution is also align with the idea of a green cloud-based architecture supporting environmental sustainability 
 \cite{alharthi2017green,alharthi2016sustainability}.
 
 The portal was implemented using (1) Ruby on Rails  programming language for the back end, which provides simplified features of PHP and it applied MVC architecture in its implementation; (2) HTML, CSS, and JavaScript as the programming language for the front end. 
 Selenium IDE was used for recording automated testing in Mozilla Firefox and Chrome Driver is used for running the test in Google Chrome.

%==================================================
\section{Core Functionality}
\label{sec:funct}

\begin{figure}[ht!]
\begin{center}
\includegraphics[scale=0.7]{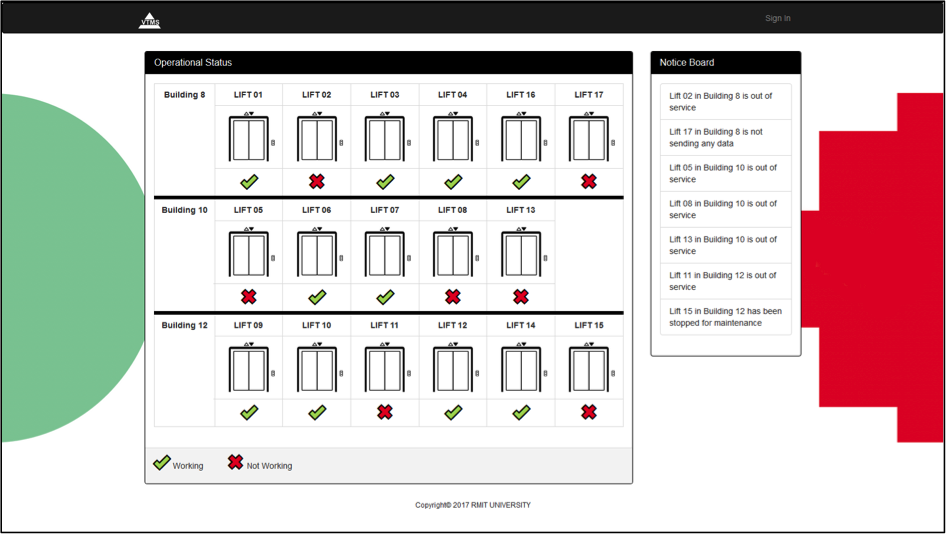}
\end{center}
\caption{Public access view}
\label{fig:public}
\end{figure}

Figure \ref{fig:public} shows the public access view of the portal. The operational status dashboard retrieves its data from database, to allow for real-time update of the status for each lift. However, the currently applied data loggers can only send the lift status with 5 minutes delay.

We use 2 symbols to describe the lift status. The green tick symbol indicates that the lift is working and the red cross symbol indicates that the lift is not working. The different shape of the symbols was selected  to make the portal accessible by helping colour-blind people to identify the lift status. The notice board also retrieves its data from database, so it is populated automatically. 
Usability and readability as well as human factors are crucial for software development, cf. \cite{spichkova2015human}, so the project aimed to take these aspects into account.

The system will send an email to the VT Management staff when the status of the lift is updated to not working to notify them about the lift status change. We use the notice board to display more information about lifts that are not working. There are 3 operation modes for lists that are not working: 
\begin{itemize}
\item out of service, 
\item no communication (the sensors are not sending any data), and 
\item in maintenance.
\end{itemize}
 There is a sign in link on the top right of the page for authorised users. Clicking the link redirects user to sign-in page.
 
\begin{figure}[ht!]
\begin{center}
\includegraphics[width=7cm]{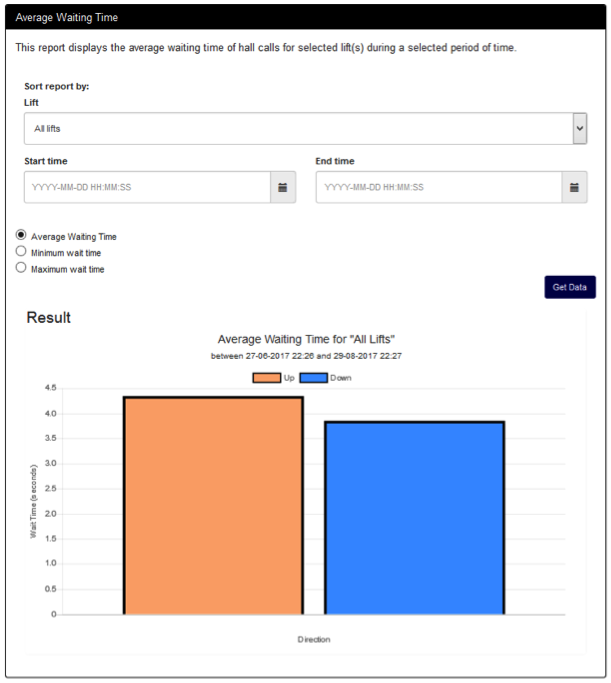}
\end{center}
\caption{Visualisation of the data on waiting times}
\label{fig:waiting}
\end{figure}

As demonstrated in Figure \ref{fig:waiting}, the system allows users to check the average waiting time page for the lifts. When the user accesses the page for the first time, it will show the average waiting time for all lifts from one day before until the day user accesses the page. The average waiting time is displayed in the form of bar chart. The data is displayed in 2 bars: up hall call and down hall call. A form is provided on the top part of the page for sorting the data based on building or lift number, start time, and end time. 

For average waiting time, the user can select if they want to view the data based on overall average waiting time, maximum waiting time, or minimum waiting time. The average waiting times can be demonstrated for a specific lift, all lifts in a building (if building number is selected), or all lifts (if the option "All Lifts" is selected) from start time until end time. If there is no data recorded between those times, then the page will display "No Data Available".

In a similar way we visualise the statistics on the 
number of hall calls page for the lifts.  
When the user accesses the page for the first time, it will show the number of hall calls for all lifts from one day before until the day user accesses the page. A form is provided on the top part of the page for sorting the data based on building or lift number, start time, and end time. Unlike average waiting time page, number of hall calls page does not have radio buttons, because there is no maximum and minimum value for number of hall calls. The data is displayed based on the accumulation of hall calls that happened between those times. When the user clicks "Get Data", the page will load the number of hall calls for a specific lift, all lifts in a building (if building number was selected), or all lifts (if "All Lifts" was selected) from start time until end time. If there is no data recorded between those times, then the page will display "No Data Available".

\begin{figure}[ht!]
\begin{center}
\includegraphics[width=7cm]{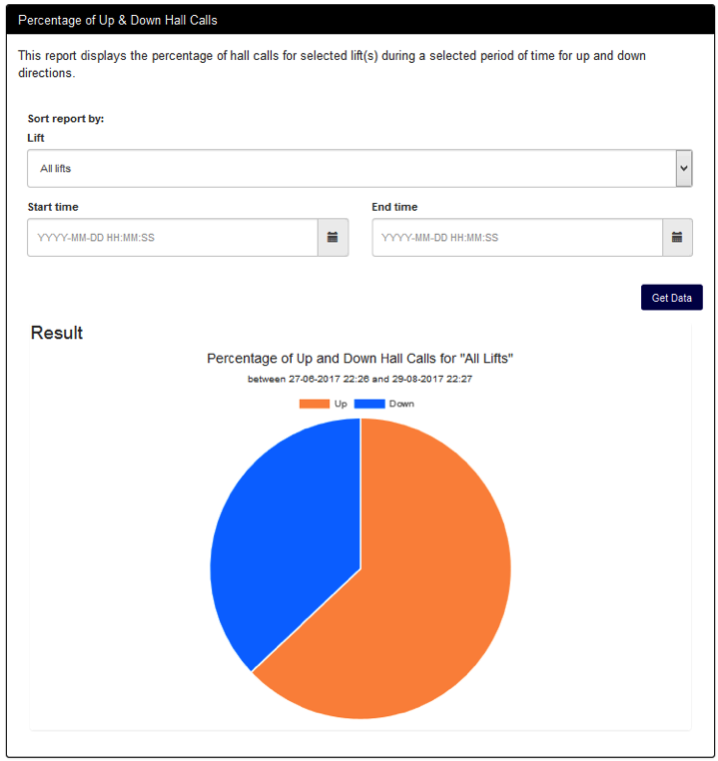}
\end{center}
\caption{Visualisation of the data on percentages of up and down hall calls}
\label{fig:perc}
\end{figure}

\begin{figure}[ht!]
\begin{center}
\includegraphics[width=6.8cm]{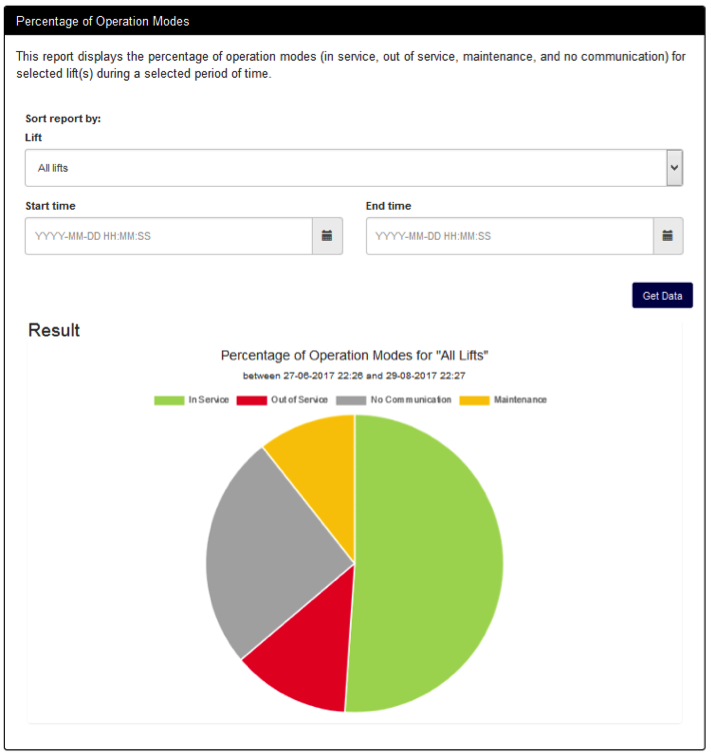}
\end{center}
\caption{Visualisation of the data on percentages of the operation modes for the lifts}
\label{fig:modes}
\end{figure}

The system also allows users to check percentages of up and down hall calls page  as well as the operation modes for the lifts, cf. Figures~\ref{fig:perc} and \ref{fig:modes}.

\begin{figure}[ht!]
\begin{center}
\includegraphics[width=8cm]{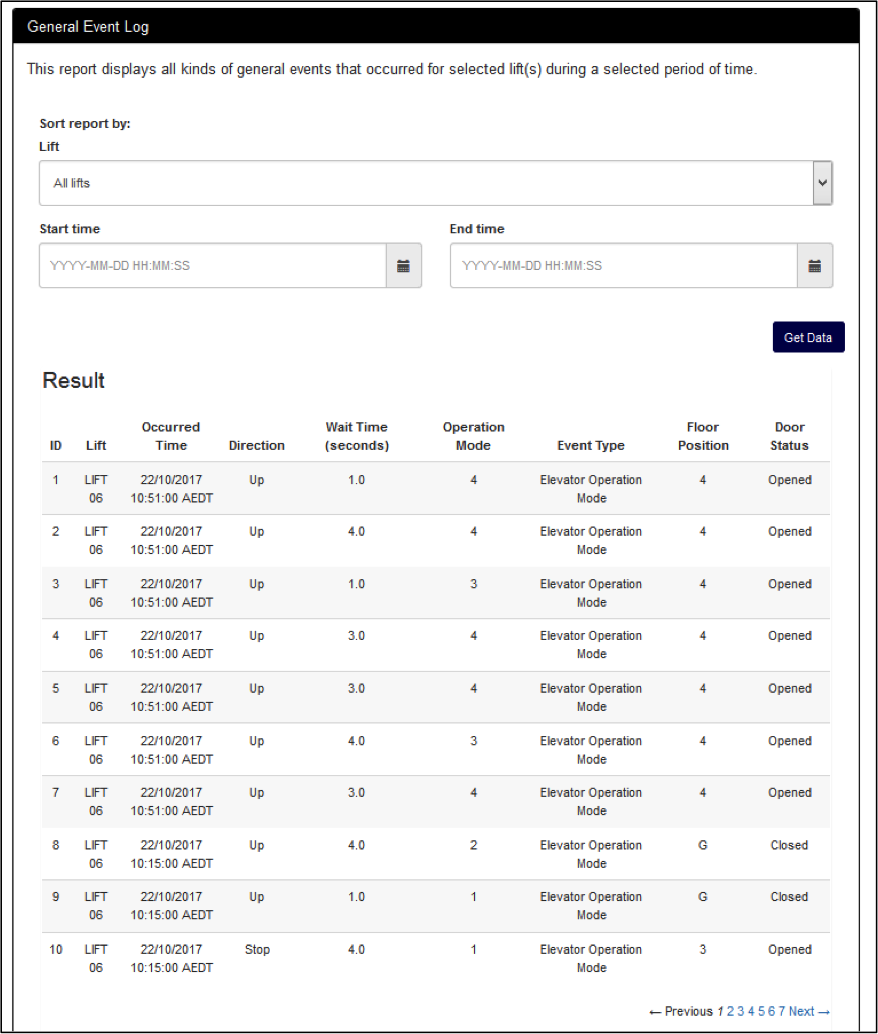}
\end{center}
\caption{Log of  general events}
\label{fig:log1}
\end{figure}

The log of general events can be checked in real time, cf. Figure \ref{fig:log1}. General event log displays all fields on every log entries: lift ID, occurred time, direction, wait time (in seconds), operation mode ID, event type, floor position, and door status. A form is provided on the top part of the page for sorting the data based on building or lift number, start time, and end time. 
The portal also provide similar representation of the logs for 
hall calls and emergency events, as well as sign-in history.

%==================================================
\section{Discussion and Conclusions}
\label{sec:conclusions}

In this paper, we presented the core results of the project on visualisation and analysis of data collected from the vertical transport facilities. 
The proposed solution was implemented as a web-based system for the 
New Academic Street (NAS) precinct of the RMIT University, Australia. The system allows the users
to monitor lifts and escalators in the NAS precinct and analyse their performances. The aim of the project was to provide better user experience for students and staffs when they used vertical transportation modes, as well as  to help building maintenance staffs doing their job in a more efficient way, resulting in increased productivity.
The core parts of the system are two web applications for vertical transportation to cover  the needs of (1) staff working on building maintenance, (1) students and university staff  who are regularly using the facilities in the corresponding buildings.
 
%==================================================
\section*{Acknowledgements} 
The project was conducted under support of New Academic Street (NAS), RMIT University.
We would like to thank Jeremy Elia,  
Belinda Kennedy, and  
Elicy Lay for numerous discussions and support.
 
\bibliographystyle{abbrv}
%\bibliography{biblio} 

\end{document}